\documentclass[nofootinbib, aps, prd, a4paper, 10pt, superscriptaddress, eqsecnum, showkeys]{revtex4-2}
\usepackage[a4paper, top=2cm, bottom=2cm, left=2.5cm, right=2.5cm]{geometry}

\usepackage{amsmath}
\usepackage{amsfonts}
\usepackage{amsthm}
\usepackage{bm}
\usepackage{mathrsfs}
\usepackage{braket}

\usepackage{graphicx}
\usepackage{booktabs}

\usepackage{xcolor}
\usepackage[pdfusetitle]{hyperref}
\hypersetup{colorlinks = true, allcolors = blue}

\usepackage{orcidlink}

\begin{document}

\title{From Quantum Correlations to Inflationary Tracking Scalar Field Evolution}
\author{V.K. Oikonomou\orcidlink{0000-0003-0125-4160}}
\email{voikonomou@gapps.auth.gr} \affiliation{Department of
Physics, Aristotle University of Thessaloniki, Thessaloniki 54124,
Greece} \affiliation{Center for Theoretical Physics, Khazar
University, 41 Mehseti Str., Baku, AZ-1096, Azerbaijan}

\begin{abstract}
The tracking condition $\dot{\phi}^2=\gamma H^{-m}$ for single
scalar field theory leads to analytic inflationary solutions,
which are compatible with the current cosmic microwave background
radiation experiments. To our knowledge this is the only analytic
solution of inflation which is compatible with the data, to date.
In this work we seek for some theoretical basis that can lead to
the tracking condition $\dot{\phi}^2=\gamma H^{-m}$. As we show,
if the Universe is seen pre-inflationary as a quantum statistical
system, the scalar field may emerge as a collective condensate of
the quantum degrees of freedom. In the quantum-to-classical
transition of the Universe, the scalar field two point function
yields the susceptibility of the theory with an inherent
correlation length. Using information theoretic motivation and
Wilsonian quantum field theoretic arguments, we demonstrate that
the tracking conditions $\dot{\phi}^2=\gamma H^{-m}$ emerge from
this framework.
\end{abstract}

\maketitle

\section{Introduction}

The origin of the Universe during its primordial genesis is a
mystery which probably will never be fully understood. For this
era one may speculate on how the quantum Universe emerged and
evolved, but there is no experiment that can actually probe the
Planck scale in order to understand the quantum degrees of freedom
of the primordial Universe, its dimension, its topology and
geometry, if such things could be defined properly primordially.
String theory succeeded in providing a possible ultraviolet (UV)
completion of the Standard Model (SM) of particle physics, but the
experimental verification of this theory is far from being a
realistic task.

The only possibility of consistently probing the effective
low-energy Lagrangian of the UV-completed theory is offered by
inflation
\cite{inflation1,inflation1a,inflation2,inflation3,inflation4}.
The inflationary regime emerged just after the quantum era of our
Universe and it is a classical regime, meaning that the spacetime
has definite geometry and topology during inflation and the
Universe is four dimensional. Since the inflationary regime is
chronologically so close to the quantum era of our Universe, it
may carry imprints of the quantum regime of our Universe, for
example in the form of low-energy string corrections in the
inflationary Lagrangian, like for example in the case of
Einstein-Gauss-Bonnet theories
\cite{Hwang:2005hb,Nojiri:2006je,Cognola:2006sp,Nojiri:2005vv,Nojiri:2005jg,Satoh:2007gn,Bamba:2014zoa,Yi:2018gse,Guo:2009uk,Guo:2010jr,Jiang:2013gza,vandeBruck:2017voa,Pozdeeva:2020apf,Vernov:2021hxo,Pozdeeva:2021iwc,Fomin:2020hfh,DeLaurentis:2015fea,Chervon:2019sey,Nozari:2017rta,Odintsov:2018zhw,Kawai:1998ab,Yi:2018dhl,vandeBruck:2016xvt,
Maeda:2011zn,Ai:2020peo,Easther:1996yd,Codello:2015mba,Oikonomou:2021kql,Oikonomou:2022xoq,
Odintsov:2020sqy,Oikonomou:2024etl,Fier:2025huc,Bajardi:2024efo,Bajardi:2025jud,Bajardi:2023gkd,Bajardi:2023byv,Bajardi:2020mdp,Bajardi:2020osh,Capozziello:2023vvr,deMartino:2020yhq,
Capozziello:2019wfi,Capozziello:2016eaz,Capozziello:2014ioa,Capozziello:2013xn}.

Unlike the quantum Universe, the inflationary Universe is
consistently constrained by the current cosmic microwave
background (CMB) experiments \cite{ACT:2025fju,ACT:2025tim} and
will further be probed by the future CMB experiments
\cite{SimonsObservatory:2019qwx} and the future gravitational wave
experiments
\cite{Hild:2010id,Baker:2019nia,Smith:2019wny,Crowder:2005nr,Smith:2016jqs,Seto:2001qf,Kawamura:2020pcg,Bull:2018lat,LISACosmologyWorkingGroup:2022jok}.
The scientific community anticipates the experimental verification
of the B-mode in the CMB polarization, which will be a smoking gun
signal for inflation.

Inflation traditionally is realized by a minimally coupled scalar
field
\cite{inflation1,inflation1a,inflation2,inflation3,inflation4},
but it can also be realized geometrically by modified gravity
theories \cite{reviews1,reviews2,reviews3,reviews4}. Recently, a
new analytic solution for inflation was developed in Ref.
\cite{Oikonomou:2026mvp}, by using the simple assumption that the
scalar field satisfies a generalized tracking condition
$\dot{\phi}^2=\gamma H^{-m}$. Tracking scalar field conditions
also appear in the literature \cite{Brax:2014wla} in dark energy
contexts. In Ref. \cite{Oikonomou:2026mvp} the tracking relation
yielded full analyticity of the Friedmann and Raychaudhuri
relations and thus the Hubble rate and the scalar field as a
functions of time were derived. The resulting inflationary theory
yielded exceptionally simple expressions for the spectral index of
the scalar perturbations and the tensor-to-scalar ratio, as
follows,
\begin{equation}\label{nslargeN}
n_{\mathcal{S}}\simeq 1-\frac{m+4}{(m+2) N}+\frac{m+4}{(m+2)^2
N^2}\, ,
\end{equation}
\begin{equation}\label{rlargeN}
r\simeq \frac{16}{(m+2) N}-\frac{16}{(m+2)^2 N^2}\, .
\end{equation}
and they solely depend on the parameter $m$ and the $e$-foldings
number. This inflationary theory is compatible with the ACT data
\cite{ACT:2025fju,ACT:2025tim}, and although in the absence of
ordinary perfect matter fluids the evolution is leading to a
pressure singularity, the inclusion of classical perfect fluids
makes the cosmological system able to avoid the pressure
singularity classically \cite{Oikonomou:2026mvp}, and even horizon
quantum phenomena may also help avoiding the singularity.

In this work we aim to provide some theoretical motivation for the
tracking condition $\dot{\phi}^2=\gamma H^{-m}$. We shall base the
discussion on the information theoretic perspective of the quantum
era of our Universe, in which the scalar field is viewed as a
collective condensate of the quantum degrees of freedom before
spacetime as we classically know it took its form. The existence
of scalar fields in cosmology is not ad hoc. In string theory
scalar fields emerge as string moduli, and there are quite many
types of scalar fields. Axions are scalar fields, which are also
string motivated in terms of ultra-light axion fields, and even
the Higgs is a scalar field. In the context of our approach the
scalar field is a statistical condensate of the quantum degrees of
freedom, just like magnetization is the statistical average of
spins in ferromagnet. This condensation effect takes place near
the quantum-to-classical transition of our Universe. This is a far
from equilibrium process, so using a Wilsonian field theoretic
basis and an information theoretic basis we connect the kinetic
energy of the scalar field to the susceptibility of the Wilsonian
theory, which in turn is related to the correlation length of the
Wilsonian two-point function. The tracking condition for the
scalar field $\dot{\phi}^2=\gamma H^{-m}$ emerges if it is assumed
that the correlation length is related to the only physical length
after the quantum-to-classical transition, the Hubble radius.

\section{The Quantum Information Origin of Scalar Condensates and Emergent Cosmological Dynamics}

Inflation will be further probed by the next stage CMB experiments
in the next few years and this will further shed light on this
primordial era of our Universe. But the inflationary regime is
essentially a classical part of our Universe, the Universe has a
definite dimension, a specific topology and geometry. The
difficult part of describing the primordial Universe is the
pre-inflationary era. That era provides mental space for fruitful
speculations which can never be verified experimentally, unless
these speculations are tight to the classical era via some
constraint on it or even via higher order gravitational terms that
may be probed, such as Einstein-Gauss-Bonnet terms
\cite{Hwang:2005hb,Nojiri:2006je,Cognola:2006sp,Nojiri:2005vv,Nojiri:2005jg,Satoh:2007gn,Bamba:2014zoa,Yi:2018gse,Guo:2009uk,Guo:2010jr,Jiang:2013gza,vandeBruck:2017voa,Pozdeeva:2020apf,Vernov:2021hxo,Pozdeeva:2021iwc,Fomin:2020hfh,DeLaurentis:2015fea,Chervon:2019sey,Nozari:2017rta,Odintsov:2018zhw,Kawai:1998ab,Yi:2018dhl,vandeBruck:2016xvt,
Maeda:2011zn,Ai:2020peo,Easther:1996yd,Codello:2015mba,Oikonomou:2021kql,Oikonomou:2022xoq,
Odintsov:2020sqy,Oikonomou:2024etl,Fier:2025huc,Bajardi:2024efo,Bajardi:2025jud,Bajardi:2023gkd,Bajardi:2023byv,Bajardi:2020mdp,Bajardi:2020osh,Capozziello:2023vvr,deMartino:2020yhq,
Capozziello:2019wfi,Capozziello:2016eaz,Capozziello:2014ioa,Capozziello:2013xn}.

As we mentioned in the introduction, in this work we shall try to
provide a theoretical motivation for the tracking condition of
scalar inflationary theories. The exact constraint will be derived
by information theoretic principles, viewing spacetime as tiles of
quantum information with correlations among them. The
pre-inflationary era belongs to the quantum gravity regime, and in
such frameworks a spacetime Hamiltonian is found from the quantum
degrees of freedom. Utilizing information theoretic approaches and
condensed matter motivated physics, we shall demonstrate that the
tracking conditions of the inflationary era may emerge naturally.

Let us go through our qualitative theoretical proposal and
gradually we analyze how the above description is realized.
Starting off with the pre-inflationary quantum gravity era, the
degrees of freedom active in that era are questionable and the
whole subject is somewhat speculative. In the classical era, the
inflaton is a spin zero scalar field which is responsible for the
cosmic acceleration of the Universe, and the inflaton is a
fundamental scalar field which must originate from the quantum
gravity era. It is thus a remnant of the quantum era, which is
classicalized when spacetime becomes four dimensional with
specific geometry and topology.

\begin{figure}
\centering
\includegraphics[width=18pc]{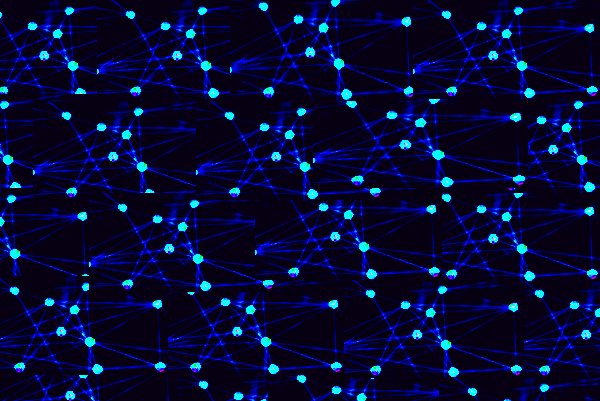}
\includegraphics[width=19pc]{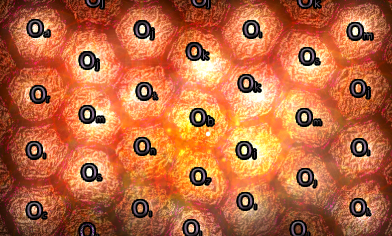}
\includegraphics[width=19pc]{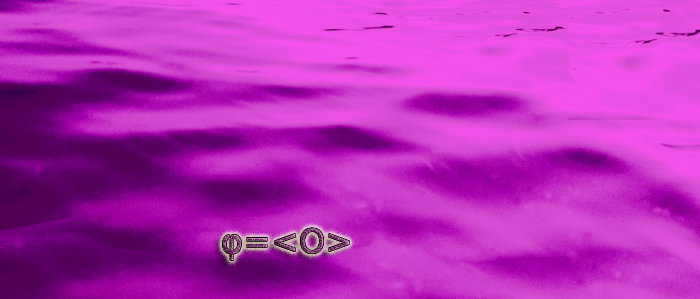}
\caption{The Universe depicted as information framework.
Primordially, quantum degrees of freedom are loosely correlated
(upper left plot), but gradually the correlations between the
quantum degrees of freedom increase (upper right plot) and
eventually, continuous spacetime and the scalar field as
information condensate of the quantum degrees of freedom emerge
(bottom plot).}\label{informationuniverse}
\end{figure}

Although this scalar field description is quite successful in
producing the desirable CMB features necessary for structure
formation, the scalar field itself is a mystery originated in the
quantum era. String theory has a large number of string moduli
which can serve as inflatons, and thus there is strong motivation
for the presence of scalar fields in the Universe and their
fundamental role. Let us not also forget the Higgs, which is a
scalar field and can also be the inflaton. In this work we aim to
present an information theoretic motivation and description of the
scalar field. So we fundamentally ask what if the scalar field
itself is a collective degree of freedom which arises from an
emergent spacetime from a quantum gravity era? Specifically, we
aim to answer this question by viewing the scalar field as a
collective condensate of quantum states near the quantum
spacetime-classical spacetime transition. Note that this
transition is a far from thermodynamic equilibrium process. In
Fig. \ref{informationuniverse} we present schematically an
artistic viewpoint of the quantum-to-classical transition and how
the continuous spacetime and the scalar field which is a
correlation condensate, may emerge by the quantum degrees of
freedom, the correlations of which increase near the transition.
This perspective of having spacetime and quantum fields emerging
from a deeper non-geometric in the classical sense quantum
substrate, is one of the main aspects that constitute modern
quantum gravity. Thus the Universe as a whole before its
classicalization is not described by particles or quantum fields
or even strings that live on a pre-existing and well defined
spacetime manifold. In the line of research we adopt in this
article, the fundamental description of the Universe involves
microscopic quantum degrees of freedom having primary
characteristics which are not momenta and positions in spacetime,
spacetime is not formed, but their characteristics are
correlations, entanglement and information content. The Universe
is thus comprised by a quantum information network primordially,
before spacetime was formed.

This description is not entirely disconnected from physical
reality, since there exist analogies in condensed matter physics.
In such contexts, a macroscopic phase of matter is not described
by the microscopic constituents themselves, but matter is
described collectively from collective variables which emerge
after coarse graining, like for example magnetization and
Bose-Einstein condensation. These condensed matter phenomena are
not fundamental microscopic objects but result from collective
manifestation of the course of organization of many underlying
degrees of freedom in the physical system, which are not analyzed
distinctively whatsoever.

An important feature of transitions in these condensed matter
phenomena, is that some microscopic disordered phase transits to
some macroscopic ordered phase the transitions are governed by
universal principles of the critical phenomena, correlation growth
and even renormalization group flows in quantum field theoretic
settings. We shall employ this perspective in the analysis that
will follow.

Turning back from the analogy of condensed matter physics to
quantum gravity Universe before classicalization, we may view the
quantum gravity regime as a physical system comprised by quantum
cells of some form, equipped with Planck degrees of freedom. These
cells are not physical cells occupying space, space did not exist
primordially as we know it, but these can be viewed by our
description as elementary quantum information units which are
connected via a network of correlations. Hence, in this quantum
gravity era perspective, the fundamental Universe structure is a
quantum information network and not classical spacetime. The
strength of the correlations and their organization will determine
the emergence of collective macroscopic variables, which may
persist after the classicalization of the Universe. In this line
of research, the scalar field itself can be a collective
excitation of quantum cells belonging to the primordial quantum
information network, just like the formation of an order parameter
in a many-body system, or the magnetization in a ferromagnet which
measures the total alignment of the spins. Hence, the scalar field
is not a particle, thus it is viewed as a collective condensate
which describes the coherent organization of the various
microscopic degrees of freedom which describe the Universe during
its quantum regime.

Now the scalar field as a collective coherence emerges near the
transition of the quantum Universe to the classical Universe
during which the inflationary era takes place. What one expects
near this transition is quite important for the analysis that
follows. The phase transitions are governed by the physics of
critical phenomena and near a phase transition, long range
correlations are developed in microscopic systems which experience
phase transitions. These long-range correlations are characterized
by a fundamental length, which is unique and typical of the length
that measures the size of the regions which fluctuate in a
coherent way. This length is a characteristic measure of the
correlated areas essentially and it is called correlation length.
In statistical physics there is an important physical variable
called susceptibility, which measures the reaction of a physical
system to external perturbations and specifically it measures the
total effect of microscopic correlations in the system as a whole.
The susceptibility grows near a critical phase transition because
the system's fluctuations become strongly correlated for larger
scales within the physical system. These phenomena classify
materials to universality classes and indeed these are collective
phenomena which are overall independent of the details of the
microscopic system.

Our approach to the quantum Universe regime before it
classicalized will be inspired by such phenomena. Specifically, we
shall assume that the quantum-to-classical transition of the
Universe is a sort of ordering transition, and the emergence of a
classical spacetime and of a scalar field is due to the
large-scale collective phenomena in the quantum Universe.
Specifically, the microscopic quantum degrees of freedom near the
transition correlate strongly and the coherence between these
quantum microscopic degrees of freedom produces spacetime and the
scalar field as a condensate state. This quantum condensate, the
scalar field, should carry important information for the
correlations of the microscopic quantum degrees of freedom.
Specifically, the correlations length and the related
susceptibility should carry critical information for the
microscopic system. Remarkably, in our proposal, such information
could be probed by CMB experiments, since the observables depend
on the parameter $m$ which is related to the critical exponents
and enters the tracking condition $\dot{\phi}^2=\gamma H^{-m}$.

The above perspective is not strange to modern quantum
gravitational theories, for example in standard black hole
thermodynamics, the area of the horizon of a black hole is
directly related to the information variable $S$, the entropy.
Hence, the horizons are systems that store information and
therefore the microscopic degrees of freedom of the black hole
determine the macroscopic properties of the black hole. The same
principle applies in holographic physics, where geometry and
spacetime geometry may be linked uniquely. In addition, modern
entropic-gravity physics research
\cite{Nojiri:2026ish,Cognola:2013fva,Carroll:2016lku,Easson:2010av}
exploits the very fact that deviations from the Bekenstein entropy
law \cite{Bekenstein:1995un,Bekenstein:1993dz} may themselves be
manifestations of modified gravity. These theories basically imply
that the distinct collective organizations of microscopic quantum
degrees of freedom lead to distinct gravities.

But the closest analogy of our perspective of quantum Universe and
its transition to its classical regime is quantum information
theory and information geometry. In this context the actual
physical quantum states are points in information space. The
physical distance between points (the quantum states) is described
by information theory variables, not geometric lengths, and the
same applies for fluctuations.

Hence, hereafter, we shall adopt the information theoretic
perspective for the quantum pre-inflationary Universe near the
critical transition. In a nutshell, geometry of spacetime is
generated by information geometry of quantum states, and the
correlations of these primordial quantum states produce a
condensate field, the scalar field, which emerges along with
spacetime continuum itself. This is schematically summarized in
Eq. (\ref{schem}), and this is our core assumption for this
article.
\begin{align}\label{schem}
\,\,\,\,\,\,\,& \mathrm{Planck}\,\,\, \mathrm{Cells}\,\,\, \mathrm{Information} \,\,\,\,\,\,\,\\ \notag & \downarrow \\
\notag &
\mathrm{Quantum}\,\,\, \mathrm{correlations} \\ \notag & \downarrow \\ \notag & \mathrm{Spacetime}\,\,\, \mathrm{Geometry}\\
\notag & \downarrow \\ \notag &
       \mathrm{Scalar}\,\,\,\mathrm{condensate}
\end{align}
Now the important part of our approach is that it is not just a
theoretical proposal, but it has important implications on the
dynamics of the scalar field and the subsequent inflationary
dynamics. Specifically, our considerations lead to a non-trivial
kinetic evolution of the scalar field the parametrization of which
is linked to the quantum correlations of the microscopic degrees
of freedom.

\section{From Quantum Correlations to Inflationary Scalar Condensates}

\subsection{Viewing the Scalar Field as a Condensate of Primordial Quantum Degrees of Freedom}

In cosmology, the existence of scalar fields is attributed to the
UV-completions of the SM. However, following the discussion of the
previous section, in this work we introduce the fundamental idea
that the scalar field is not an existing particle, but it is a
statistical condensate of quantum degrees of freedom, so it is a
macroscopic effect of microscopic degrees of freedom at their
maximum correlation. The correlation length is something that
should be propagated in the scalar field. Let us demonstrate how
such a behavior may be realized in nature.

Let us start with an arbitrary microscopic Hamiltonian describing
the quantum degrees of freedom $\hat{O}_i$,
\begin{equation}\label{ref}
\hat{H} = \hat{H}_0 - \sum_{ij} J_{ij}\, \hat{O}_i \hat{O}_j \, ,
\end{equation}
and we define the two-point correlation function as follows,
\begin{equation}\label{ref1}
G_{ij} = \langle \hat{O}_i \hat{O}_j \rangle - \langle \hat{O}_i
\rangle \langle \hat{O}_j \rangle \, .
\end{equation}
Then, according to the discussion of the previous section, the
scalar field is a collective emergent condensate of the quantum
degrees of freedom near the critical quantum-to-classical
transition,
\begin{equation}\label{ref2}
\phi(\mathbf{x}) = \langle \hat{O}(\mathbf{x}) \rangle \, .
\end{equation}
The above considerations are essentially basic statistical
mechanics, in which one may view the scalar field and the
classical spacetime as organizational manifestations of the
microscopic quantum degrees of freedom $\hat{O}_i$ to a condensate
collective state. This proposed perspective aligns with
ferromagnets in which the magnetization is the average of
microscopic spin degrees of freedom  $M=\langle S\rangle $, or
even like Bose-Einstein condensation.

\subsection{Coarse-grained  Dynamics of the Condensate Scalar Field}

Now, let us give some dynamics in the condensate scalar field.
Recall that the scalar field as a collective condensate of
microscopic degrees of freedom emerges near criticality, at the
quantum-to-classical transition. This transition is not an
equilibrium process, but a far off equilibrium process, so
thermodynamical equilibrium of any sort does not apply. But since
the scalar field is basically a function of the newly four
dimensional emerged spacetime $\phi (x^{\mu})$, it will belong to
a Wilsonian field theoretic framework. The lowest-order effective
action which governs the long-wavelength degrees of freedom is
$F(\phi)$, with functional,
\begin{equation}\label{wilson1}
Z = \int \mathcal{D}\phi\, e^{-F[\phi]} \,,
\end{equation}
and a general analytic form,
\begin{equation}\label{wilson2}
F[\phi] = \int d^d x \left[ \frac{K}{2} (\nabla\phi)^2 + V(\phi)
\right]\, ,
\end{equation}
and we keep the dimension of the emergent spacetime arbitrary $d$,
but in the end we must choose $d=4$. Standard maximization process
of the action implies,
\begin{equation}\label{wil3}
\frac{\delta F}{\delta\phi}=0\, .
\end{equation}
The above maximization principle does not follow from equilibrium
thermodynamics principle, it is simple Wilsonian field theory.
Suppose that the solution of the maximization is a scalar field
value $\bar{\phi}$, so we expand the scalar field around this
background solution,
\begin{equation}\label{wil4}
\phi=\bar{\phi}+\delta\phi,
\end{equation}
and we get,
\begin{equation}\label{wil5}
F[\phi] = F[\bar{\phi}] + F'(\bar{\phi})\,\delta\phi + \frac{1}{2}
F''(\bar{\phi}) (\delta\phi)^2 +... \, .
\end{equation}
Since $\bar{\phi}$ satisfies,
\begin{equation}\label{wil6}
F'(\bar{\phi})=0\, .
\end{equation}
the quadratic coefficient $F''(\bar{\phi})$ essentially defines
the inverse response kernel of the Wilsonian (coarse-grained)
theory,
\begin{equation}\label{wil7}
\Gamma^{(2)}(x,y) = \left. \frac{\delta^2F}
{\delta\phi(x)\delta\phi(y)} \right|_{\phi=\bar{\phi}} \, .
\end{equation}
and the corresponding two-point function is,
\begin{equation}\label{wil8}
G(x,y)=\left[\Gamma^{(2)}(x,y)\right]^{-1}\, .
\end{equation}
Now the above two-point function is related to the susceptibility
of the coarse-grained theory, defined as follows,
\begin{equation}\label{wil9}
\chi = G(k=0) = \int d^dy\, G(x,y)\, .
\end{equation}
Hence, the susceptibility naturally emerges in the Wilsonian
framework, as the inverse curvature of the functional which
governs the long wavelengths of the coarse grained theory.

Considering the two-point function near criticality, at the fixed
point of the theory, the two-point function has the following
universal form,
\begin{equation}\label{uni1}
G(r) = \frac{1}{r^{d-2+\eta}} f\!\left(\frac{r}{\xi}\right),
\end{equation}
with $\eta$ being the anomalous dimension, $\xi$ being the
correlation length at criticality and we keep the emergent
spacetime's dimension arbitrary. Recall the definition of the
susceptibility function $\chi$ in Eq. (\ref{wil9}), which in terms
of the radial coordinate is written,
\begin{equation}\label{uni2}
\chi = \int d^dr\,G(r),
\end{equation}
so by integrating using (\ref{uni1}), we get the following
universal scaling,
\begin{equation}\label{uni3}
\chi \sim \xi^{2-\eta} \, .
\end{equation}
Now we need to make a critical assumption for the analysis, which
is however physically motivated from dimensional reasons. From
dimensional arguments, the correlation length must be related to a
fundamental length of the newly emerged spacetime. The only such
length is the Hubble radius,
\begin{equation}\label{h1}
R_H=\frac{1}{H}\, ,
\end{equation}
thus we generally assume that the correlation length is related to
the Hubble radius as follows,
\begin{equation}\label{h2}
\xi = \xi_0 R_H^{z}=\xi_0 H^{-z}\, ,
\end{equation}
with $z>0$ some arbitrary critical exponent which quantifies
essentially the critical transition of the quantum-to-classical
transition. Hence, in view of the relation (\ref{h2}), the
susceptibility is related to the Hubble rate as follows,
\begin{equation}\label{h3}
\chi \sim \xi^{\,2-\eta} \sim H^{-z(2-\eta)}\, .
\end{equation}

Now coming to the evolution of the scalar field, it satisfies,
\begin{equation}\label{h4}
\dot{\phi} = \frac{\delta\phi}{\delta\rho}\, \dot{\rho}\, ,
\end{equation}
where $\rho$ is the microscopic density matrix, so the above
relation implies,
\begin{equation}\label{h5}
\frac{1}{2}\dot{\phi}^{\,2} = \frac12 \left(
\frac{\delta\phi}{\delta\rho} \right)^2 \dot{\rho}^{\,2} \, .
\end{equation}
The above equation (\ref{h5}) can be viewed in an information
theoretic basis, because the microscopic evolution is embodied in
the time dependence of the microscopic density matrix and
specifically on $\dot{\rho}^{\,2}$. The latter measures the rate
at which the microscopic degrees of freedom are organized near the
quantum-to-classical transition. We can use an information
geometric framework to embed the above quantities. Specifically,
the squared norm of the tangent vector $\dot{\rho}$ defines a
metric on the manifold of the quantum states. This measures how
the quantum states are infinitesimally distinguished on the
manifold of quantum states. The integrated information measure is
defined as,
\begin{equation}\label{h6new}
\mathcal{I} \equiv \|\dot{\rho}\|^2\, .
\end{equation}
The exact metric will depend on the full microscopic information
of the quantum system, for example Fisher information, but in our
case we merely need its scaling properties. Equation (\ref{h5})
therefore becomes,
\begin{equation}\label{h7new}
\frac12\dot{\phi}^{\,2} = \alpha\,\mathcal{I}\, ,
\end{equation}
where the constant $\alpha$ contains the coarse-graining factor
$\frac12(\delta\phi/\delta\rho)^2$. The information variable
$\mathcal{I}$ essentially quantifies the total amount of the
microscopic coherence and the simplest realization is if it
relates to the susceptibility, The simplest realization is
\begin{equation}\label{h8}
\mathcal{I} \propto \chi \, ,
\end{equation}
which is another quantity that measures the exact same thing, the
microscopic coherence. Therefore in view of Eqs. (\ref{h3}),
(\ref{h7new}) and (\ref{h8}), one easily gets,
\begin{equation}\label{h9}
\dot{\phi}^{\,2} = \gamma H^{-z(2-\eta)}\, .
\end{equation}
Upon defining,
\begin{equation}\label{h10}
m = z(2-\eta)\, ,
\end{equation}
we obtain,
\begin{equation}\label{h11}
\dot{\phi}^{\,2} = \gamma H^{-m} \, ,
\end{equation}
which is the tracking condition used in Ref.
\cite{Oikonomou:2026mvp}. The tracking exponent $m=z(2-\eta)$
contains the anomalous dimension $\eta$ and also the critical
exponent $z$, which describes they way that correlations scale at
the transition point of the quantum-to-classical transition of the
Universe. Therefore, the tracking condition may be viewed as
having a microscopic critical phenomena based interpretation,
rather being a purely phenomenological ansatz.

Note that, if instead assuming that the transition from quantum to
a classical Universe is viewed as a critical phase transition, and
it is viewed as an equilibrium process, then using the equilibrium
Landau-Khalatnikov equation,
\begin{equation}\label{h14}
\dot{\phi} = - \Gamma \frac{\delta F}{\delta\phi}\, ,
\end{equation}
after some simple considerations we would obtain the following
scaling,
\begin{equation}\label{h15}
\dot{\phi}^{\,2} \sim \chi^{-1} \sim \xi^{-(2-\eta)} \, ,
\end{equation}
which quite different from what Eq. (\ref{uni3}). But the
classicalization transition is a far from equilibrium process.

\section{Conclusions}

In this work we provided a qualitative motivation for the tracking
condition $\dot{\phi}^2=\gamma H^{-m}$ for the scalar field
inflation. Such tracking condition is known to lead to analytic
inflationary solutions which in turn lead to viable inflationary
phenomenology \cite{Oikonomou:2026mvp}. The whole analysis is
based on the perspective that the scalar field is an information
theoretic condensate of the quantum degrees of freedom of the
Universe prior its classicalization. Utilizing Wilsonian field
theoretic effects and critical phenomena, we connected the kinetic
energy of the scalar field to the susceptibility, which in turn is
related to a the correlation length of the collective excitations
of the quantum degrees of freedom during the classicalization
process. The analysis we performed is merely qualitative and a
more detailed microscopic theory is required in order for it to be
fundamentally based. However our aim was to provide a theoretical
basis and some qualitative arguments for tracking conditions of
scalar fields. The tracking condition itself has the slow-roll
condition as an inherent characteristic, although the potential of
the scalar field can be determined in this analytic inflation
theory and it is unique. Nevertheless, no analytic viable
inflation solutions for inflation exist, apart from the analytic
solution obtained by assuming the tracking condition
$\dot{\phi}^2=\gamma H^{-m}$, so there might be a deeper reason
for the existence of the tracking condition. In this article we
outlined how such a theoretical motivation may be constructed.

\end{document}